# Shape resonance induced electron attachment to Cytosine: The effect of aqueous media

Pooja Verma[1], Madhubani Mukherjee[2] and Achintya Kumar Dutta[1,*]

[1]*Department of Chemistry, Indian Institute of Technology Bombay, Powai, Mumbai-400076, India*

[2]*Department of Chemistry, University of Southern California, Los Angeles, California,90089-0482, United States*

**Abstract**

We have investigated the impact of microsolvation on the shape resonance states of nucleobases, taking cytosine as a case study. To characterize the resonance position and decay width of the metastable states, we employed the newly developed DLPNO-based EA-EOM-CCSD method in conjunction with resonance via Padé (RVP) approximation. Our calculations show that the presence of water molecules causes a redshift in the resonance position and an increase in the lifetime for all the three lowest-lying resonance states of cytosine. Furthermore, the lowest resonance state in isolated cytosine was converted to a bound state in the presence of an aqueous environment. The obtained results are extremely sensitive to the basis set used for the calculations.

*achintya@chem.iitb.ac.in

# 1. Introduction

Extensive research has been conducted over the years, both theoretically and experimentally, to investigate the impact of ionizing radiation on chemical stability of DNA.[1–6] These studies have played a crucial role in assessing the risks associated with radiation exposure and improving the efficacy of radiation therapy for treating severe illnesses, including cancer. Ionizing radiation can interact with DNA in both direct and indirect manner,[7–10], along with a quasi-direct mechanism in which a hole is transferred from solvent to the DNA strand. These three pathways are graphically illustrated in Figure 1. A comprehensive description of the role of these pathways in DNA damage can be found in reference.[11] The secondary electrons generated by the interaction of ionizing radiation with DNA and the surrounding environment have a significant contribution in indirect damage.[12–21] These electrons transfer their kinetic energy through inelastic collisions with the surrounding solvent molecules and generate low-energy electrons (LEEs) with an energy range of 0-20 eV, which play a major role in radiation-induced DNA damage.[22–27]

There have been several attempts to identify the precise mechanism of the initial electron attachment to the DNA. Resonance-based[12,15,17,28–35] route has been acknowledged as the principal mechanism of LEE-induced DNA strand breaks in the gas phase. Here, the incoming electron occupies an unoccupied molecular orbital of the parent neutral molecule forming transient negative ion. These quasi-bound anionic states emerging from the interaction between electron and molecule survive for a long time (femtoseconds to picoseconds),[36] and the fate of the system is governed by its lifetime. It can lead to bond breaking via the dissociative electron attachment (DEA) process or undergo auto-detachment. In DEA, the anion $(M^-)^*$ gets stabilized by bond breaking to create neutral and anion fragments. Numerous studies on nucleobases have proposed that electron attachment to the nucleobase is followed by electron transfer to the backbone of the sugar-phosphate molecule, which ultimately results in the fission of the C-O bond.[13,36–46] However, evidence for the direct attachment of electrons to the sugar-phosphate bond also exists.[16,17,38,47] Generally, a transient negative ion (TNI) with a lifetime of half or more the vibrational period of the cleaved bond decays through DEA.[48] However, the TNI with a shorter lifetime undergoes auto-detachment via emission of electron from the anion (forming $M^*$ and $e^-$). The electron released may get attached to other sites of DNA, causing further damage. The TNI might alternatively release the extra energy into the environment by relaxing back to the ground state radical anion.[36]

Quasi-bound metastable states formed in nucleobases can be categorized into two groups, [29,45,49] based on the energy of the incoming electron; shape resonance and Feshbach resonance. Shape resonances are produced when an electron with an energy of 0-4 eV enters the previously unoccupied π* orbitals of the nucleobase ground state. Shape resonances may also get formed if the electron is attached to a σ* orbital, but these resonances are likely to be short lived.[29] Due to their short life span, these σ* type shape resonances are difficult to detect experimentally. However, theoretical calculations on the σ* type shape resonances of DNA subunits have been reported in the literature.[50–52] In addition to shape resonances, there are core-excited/Feshbach resonances, where an electron with energy above 4 eV is captured by the neutral excited state, involving two electrons in a virtual orbital and a hole in the occupied orbitals. If the resonance energy of the anion is greater than the corresponding neutral excited state, it is a core-excited shape resonance. However, if it is lower, then it is a Feshbach resonance. Figure 2 depicts the orbital level diagram corresponding to shape and core-excited/Feshbach resonances.

The simulation of resonance states is challenging compared to the bound states, as these states exhibit exponential growth in the asymptotic region where the excess electrons are weakly coupled to the continuum.[49] This characteristic causes the wave function describing these states to lose its square integrability, making the application of standard bound state electronic structure methods impractical. Resonance states are solutions of the time-dependent Schrodinger equation that constitute a high density of states in the continuum. As a result, they cannot be characterized by a single eigenstate in Hermitian quantum mechanics.[49] Instead, they are nonstationary solutions and belong to the continuum domain of the Hamiltonian. Nevertheless, except for their asymptotic tendency, Siegert wave functions are relatively regular for resonance states and behave like a bound state in the interaction region. Hence, the Siegert equation[53] can be used to characterize the resonance wave-function as discrete states with complex energy:

$$\Psi(x,t) = e^{(-itE_{res})}\Phi_R(x) = e^{(-t\Gamma/2)}e^{(-itE_R)}\Phi_R(x) \tag{1}$$

The function $\Phi_R(x)$ behaves similar to a bound state wave-function in the interaction region (near the nuclei), where the electron is more tightly bound to the atom/molecule. The $E_R$ and $\Gamma$ are the real and complex parts of the complex energy, $E_{res} = E_R - \Gamma/2$ and determines the resonance position (measurable energy of a given resonance) and the width, respectively.

Width ($\Gamma$) and lifetime ($\tau$) are inversely proportional ($\tau = \hbar/\Gamma$) to each other, hence, larger width corresponds to a shorter lifetime for the attached electron.

There are various techniques described in the literature to compute the complex Siegert energy, which includes complex basis functions,[54,55] analytical continuation of the matrix elements of the Hamiltonian[49] and adding a complex absorption potential (CAP)[53,56–61] to the Hamiltonian. Nonetheless, implementing these approaches involves significant modification of existing quantum chemistry software. In the present study, we have employed a rather simpler approach known as the stabilization technique,[62–64] which can use the standard electronic structure packages without any additional modification. Landau and co-workers have recently proposed resonance via Padé (RVP) approximation[65–67] as an efficient way to calculate the resonance position and decay width by analytically carrying over the data acquired in real space into the complex plane.

Additionally, the accuracy of the theoretical method used is an important factor in describing anionic resonance states. Among the various electronic structure methods available for studying resonances, the equation of motion coupled cluster (EOM-CC) method[68–71] has gained popularity due to its systematically improvable nature and the ability to access multiple states of the anion in a single calculation. However, its use is limited to small molecules due to its high computational cost and large storage requirements. To reduce the computational cost, approximations must be made to the standard equation of motion coupled cluster singles doubles (EOM-CCSD) method. Recently, Dutta *et al.* described a domain-based local pair natural orbital (DLPNO)[72] based implementation of EOM-CCSD methods (EOM-DLPNO-CCSD) for the electron affinity (EA)[73] and ionization potential (IP)[74] of large molecules. It has been demonstrated that the DLPNO-based approach offers a balance between accuracy and low computational cost, making it a suitable choice for studying electron attachment-induced properties of genetic materials[75] and can be applied to larger DNA model systems.[76] However, the application of the EA-EOM-DLPNO-CCSD method has thus far been limited to bound-state anions only.

There is a limited number of experimental studies available on the resonance states of DNA model systems in the condensed phase.[77,78] A recent experimental study on microhydrated uracil using two-dimensional photoelectron spectroscopy has been reported by Verlet et al.[77] The majority of theoretical studies on resonance states in DNA model systems have also been limited to the gas phase,[29,30,79–81] with only a few performed in aqueous media.[32,77,82] However, simulating resonance states in the aqueous environment is crucial for understanding electron attachment to biological DNA. Previous studies have demonstrated the stabilizing effect of aqueous media on shape resonances states of DNA nucleobases.[32,82] However, the electronic structure methods utilized in these investigations were not always optimal in terms of accuracy. Furthermore, the use of small basis sets has been another persistent issue. The shape resonance states of DNA nucleobases are highly sensitive to basis set size,[83] and it is currently not viable to perform shape resonance calculation using standard EOM-CCSD method and extended basis set, beyond isolated nucleobases.

This manuscript aims to investigate the effect of the aqueous environment on shape resonance states of DNA nucleobases, using the lower scaling EOM-DLPNO-CCSD method and extended basis sets, with microhydrated cytosine used as a model system.

## 2. Methods
### 2.1. EA-EOM-CCSD

The equation of motion approach is one of the simplest ways to extend the ground state coupled cluster to an excited state. In the equation of motion coupled cluster (EOM-CC) method,[68–71,84,85] the target state is generated by the action of a linear excitation operator $\hat{R}_k$ on the reference state wave-function as

$$\left|\Psi_k\right\rangle = \hat{R}_k \left|\Psi_0\right\rangle \tag{2}$$

Where $\Psi_0$ is the single reference ground state coupled cluster wave-function. It is generated by the application of an exponential operator on the zeroth order reference state, $\Phi_0$ as

$$\left|\Psi_0\right\rangle = e^{\hat{T}}\left|\Phi_0\right\rangle \tag{3}$$

Here, $\Phi_0$ is generally, but not necessarily, a Hartree-Fock single determinant and $\hat{T}$ is the coupled cluster ground state excitation operator given as:

$$\hat{T} = \sum_{i,a} t_i^a \left\{\hat{a}_a^\dagger \hat{a}_i\right\} + \sum_{i<j,a<b} t_{ij}^{ab} \left\{\hat{a}_a^\dagger \hat{a}_b^\dagger \hat{a}_j \hat{a}_i\right\} + \cdots \tag{4}$$

Where the $\hat{a}$ and $\hat{a}^\dagger$ denotes the creation and annihilation operator with respect to the Hartree-Fock determinant. The indices $i, j, \ldots$ refer to occupied or internal molecular orbitals (MO), and indices *a, b, …* correspond to virtual or external ones. The suitably chosen $\hat{R}_k$ operator can generate electron attached,[71,85] ionized,[86] and excited states[68,87] by acting on the reference state. The target state energies are obtained by diagonalizing the coupled cluster similarity transformed Hamiltonian in the basis of suitably chosen configurations.

$$\bar{H} = e^{-\hat{T}} H e^{\hat{T}} \tag{5}$$

The diagonalization of similarity transformed Hamiltonian ($\bar{H}$) within the (N+1) electron space leads to the EA-EOM-CC method.

The Schrödinger equation for the target state is given as

$$\bar{H}\hat{R}_k |\Phi_0\rangle = E_k \hat{R}_k |\Phi_0\rangle \tag{6}$$

where, $E_k$ is the energy of k$^{th}$ electron attached state. The form of the linear excitation operator, $\hat{R}_k$ for electron attached state is defined as

$$\hat{R}_K^{EA} = \sum_a r^a \hat{a}_a^\dagger + \frac{1}{2} \sum_{ab} \sum_i r_i^{ab} \hat{a}_a^\dagger \hat{a}_b^\dagger \hat{a}_i + ... \tag{7}$$

The direct energy difference between the target and the initial state, $\omega_k$ (electron affinity in the present case) can be obtained using a commutator form of the equation (6) and is shown as

$$\left[\bar{H}, \hat{R}_k\right]|\Phi_0\rangle = \omega_k \hat{R}_k |\Phi_0\rangle \tag{8}$$

The EA-EOM-CC method is generally used in the singles and doubles approximation (EA-EOM-CCSD), which scales as the O($N^6$) power of the basis set dimension (the EOM part scales as the O($N^5$)) and has a similar storage requirement as that of the ground state coupled cluster method.

## 2.2 DLPNO formulation of EA-EOM-CCSD

In the domain-based local pair natural orbital (DLPNO) formulation,[72] the occupied space is expanded using localized molecular orbitals (LMO)[88,89] while the virtual space is spanned in terms of projected atomic orbitals (PAO).[88,89] Subsequently, the PAOs are spanned in terms of pair natural orbitals (PNO).[72,90–92] Due to the short-range nature of dynamic correlation, the use of the DLPNO framework significantly reduces the cost of correlation calculation on large molecules.

In this formalism, once the Hartree-Fock equation is solved, the standard localization technique is employed to generate LMO for the occupied orbitals from atomic orbitals (AO). This is achieved using the following transformation

$$\left|\Phi_i^{LMO}\right\rangle = \sum_\mu \left|\chi_\mu\right\rangle L_{\mu i} \tag{9}$$

where, $L_{\mu i}$ represents the transformation matrices from AO to LMO, i and µ denotes the indices for LMO and AO, respectively. The virtual orbitals are expanded in terms of PAOs[89] as

$$\left|\tilde{\mu}\right\rangle = \sum_\mu \tilde{P}_{\mu\tilde{\mu}} \left|\mu\right\rangle \tag{10}$$

The PAO coefficient matrix; $\tilde{P}$ is defined as

$$P = 1 - L^{occ} L^{occ\dagger} S \tag{11}$$

The S in above equation denotes the atomic overlap matrix. The PAOs are obtained by projecting out the occupied orbital space from the atomic orbitals. The correlation domains of the PAOs for each LMO are obtained using the sparse map $L(i \to \tilde{\mu}_A)$, where $\mu_A$ represents a set of PAOs with a value above a predefined threshold (TCutDo) of differential overlap integrals (DOI) between a particular occupied LMO (i) and the PAOs in the virtual space.

$$DOI_{i\tilde{\mu}} = \sqrt{\int \left|i(r)\right|^2 \left|\tilde{\mu}(r)\right|^2} \tag{12}$$

To generate the pair domains for a pair of occupied orbitals ij, the individual domains of i and j orbitals are combined by taking their union. The use of the resolution of identity (RI) approximation for integral transformation requires the generation of fitting domains for the auxiliary basis functions of each occupied orbital, which is determined by the TCutMKN threshold. Following the generation of localized occupied and virtual orbitals, a multipole estimate of the pair correlation energy is used to perform a pre-screening. The pair energies of the surviving pairs are estimated at the semi-canonical (SC) LMP2 level of theory, and they are categorized into strong pairs, which make a large contribution to the correlation energy, and weak pairs, with a smaller contribution. The selection of weak and strong pairs is controlled by the threshold TCutPairs. The weak pairs are treated using the MP2 method and their contribution is added to the total correlation energy. For each strong pair, PNOs are generated by diagonalizing the pair density, $D_{ab}^{ij}$ for the ij pair.

$$D_{ab}^{ij} = \sum_{\tilde{a}_{ij}} d_{a\tilde{a}_{ij}} n_{\tilde{a}_{ij}} d_{\tilde{a}_{ij}b} \qquad (13)$$

The occupation number of the PNOs for pair ij is given by $n_{\tilde{a}_{ij}}$ and $d_{\tilde{a}_{ij}b}$ represents the transformation matrix between PAOs ($\tilde{\mu}$) and PNOs corresponding to pair ij. The TCutPNO threshold is used to exclude PNOs with occupation numbers below the threshold. The CCSD problem is solved for the strong pairs in a more compact virtual space provided by the truncated PNOs. The singles amplitudes are expanded in terms of 'singles PNOs', which are equivalent to the diagonal 'doubles PNOs' but are truncated using a tighter threshold. After solving the coupled cluster ground state equations, the $\bar{H}$ intermediates are generated using the ground state pair list and PNO integrals. For the strong pairs, the EOM-CCSD equations are solved in the PNO basis. The EOM singles operator ($\hat{R}_1$) are kept untruncated and the doubles operators ($\hat{R}_2$) are truncated using the singles PNOs.

$$\hat{R}_k = \sum_a r^a \hat{a}_a^\dagger + \frac{1}{2} \sum_{\tilde{a},\tilde{b},i} r_i^{\tilde{a}\tilde{b}} \hat{a}_a^\dagger \hat{a}_b^\dagger \hat{a}_i \qquad (14)$$

The $\tilde{a}$ and $\tilde{b}$ in the above equation denotes that they are in the singles PNO basis of the i$^{th}$ orbital.

Hence, the accuracy of the EA-EOM-DLPNO-CCSD electron affinity values[73] is controlled by four cutoff parameters, namely TCutDO, TCutMKN, TCutPNO, and TCutPair. Neese[93] and co-workers have developed a composite truncation scheme by controlling all four truncation parameters with a single keyword. These keywords are named LOOSEPNO, NORMALPNO, and TIGHTPNO, and they give progressively more accurate results at the expense of higher computation costs. Table 2 gives the values of individual cutoff parameters used for the three composite keywords mentioned. A detailed description of the EA-EOM-DLPNO-CCSD method, including working equations, can be found in reference.[73]

## 2.3 Resonance via Padé (RVP) Method

The RVP (Resonance via Padé) approximation[64–66] has been used to calculate the resonance energy and decay rate (width) of various resonance states. This approach works by analytically continuing Hermitian electronic structure solutions into the non-Hermitian regime using Padé approximants. It uses stabilized part of the stabilization graph between two avoided crossings, as an input and estimates the corresponding energy and width of the particular resonance state. While using the RVP method, initially, stabilization plots are generated using the standard Hermitian electronic structure method.[62–64] These graphs are produced by scaling a finite Gaussian basis set, where, exponents of the most diffuse functions are scaled, introducing a real scaling parameter alpha ($\alpha$), which at higher values, represents highly diffuse orbitals and contracted orbitals at lower values of $\alpha$ (~ 0.1). This eigenvalue spectrum (energies) as a function of $\alpha$ represents the stabilization graph. The utilization of the stabilization method aims to differentiate between resonances and continuum states, as these two exhibit distinct behaviors upon scaling, primarily due to the different characteristics of their associated wave functions.

As the value of $\alpha$ is increased continuously, the quasi-continuum spectra's discrete energy levels are significantly impacted, with some being lowered and others being raised. This is because these levels belong to the delocalized wave function and are strongly dependent on the scaling parameter. However, states localized around the resonance energy are stable as resonance states exhibit a bound-state behaviour in the interaction region, and their eigenvalues are independent of the scaling parameter. This leads to the formation of energy level crossings when the energies are plotted as a function of $\alpha$. The real resonance solutions correspond to those where the energy is invariant with respect to $\alpha$, indicating a stable region in the stabilization graph. The energy's stable part corresponds to a single function, localized in the interaction region, whereas the avoided crossings are identified by the mixing of a localized resonance function and a delocalized quasi-continuous function.[67] As a result, the stable part and avoided crossings are described as having locally analytic and non-analytic behavior, respectively. In this context, the stable part possess the ability to replicate the whole stabilization graph and contains all the essential information for performing analytical continuation into the complex plane.

The analytic continuation of real data obtained from the stable region into complex plane is done using Padé approximant by fitting the stabilized region to a ratio of two polynomials as a function of real scaling parameter $\alpha$ as

$$E(\alpha) = \frac{P(\alpha)}{Q(\alpha)} \tag{15}$$

The numerical expression of this ratio is generated using Schlessinger point method.[94] This method requires M data points from the stable region along with the corresponding energy values and truncated continued fraction is represented as

$$C_M(\alpha) = \cfrac{E(\alpha_1)}{1 + \cfrac{z_1(\alpha - \alpha_1)}{1 + \cfrac{z_2(\alpha - \alpha_2)}{\vdots\, z_{M-1}(\alpha - \alpha_{M-1})}}} \tag{16}$$

Where $z_i$ coefficients are determined recursively to satisfy the following equation

$$C_M(\alpha_i) = E(\alpha_i) \qquad i = 1, 2, \ldots, M. \tag{17}$$

Resonances are identified as stationary points in the complex plane using a complex parameter, $\eta$ to substitute $\alpha$ in fitted energy functional such that $\eta = \alpha e^{i\theta}$. Stationary points are obtained by minimizing the energy with respect to $\eta$ as

$$\left.\frac{\partial E(\eta)}{\partial \eta}\right|_{\eta^{SP}} = 0 \tag{18}$$

A statistical approach[65] is used to test the stability of the derived complex energy values with respect to small variations in the input data set. Due to a derivative of a ratio of two polynomials, a large number of stationary points are obtained. To distinguish between physical stationary points (SPs) and numerical errors, a clusterization approach was employed.[65] This method groups together a set of results that share a common energy and decay rate, forming a cluster. The final reported result is the average of the values in the cluster, while the standard deviation is calculated using various points within the cluster. Detailed description on the use of stabilization and clusterization techniques in RVP method can be found elsewhere.[63,65]

## 2.4 Complex Absorption Potential (CAP) Method

In the CAP method, a complex absorbing potential is added to the molecular Hamiltonian to absorb the diverging tail of the resonance wavefunction as[53]

$$H(\eta) = H_0 - i\eta W(r) \tag{19}$$

where $\eta$ determines the strength of the potential and W is the form of the potential. The potential W can have different forms eg. cuboid, spherical, Voronoi. The smooth Voronoi potential wraps around the molecule at a specific cutoff value ($r_{cut}$).[95,96] We get the resonance energy as

$$E_{res} = E_R - \frac{i\Gamma}{2} \tag{20}$$

Here, $E_R$ and $\Gamma$ are the resonance position and width, respectively. To minimize the perturbation due to finite $\eta$ value in the finite basis set, we calculate the $\eta$ trajectory and find the minima of $\left|\eta \frac{\partial E}{\partial \eta}\right|$. We calculate the resonance position and width at the optimal $\eta$ value, where the $\eta$ dependence is the lowest. One can further reduce the perturbation by calculating the first-order corrected energy as[97]

$$U(\eta) = E(\eta) - \left|\eta \frac{\partial E}{\partial \eta}\right| \tag{21}$$

The resonance position and width can be found by minimizing the first-order corrected trajectory with respect to $\eta$. To minimize the computational cost of the calculations, we have used the projected CAP-EOM-CCSD scheme of Gayvert and Bravya.[98] In this approach, the CAP-augmented Hamiltonian is represented at a subspace consisting of a small number of real eigenstates of the EOM-CC Hamiltonian, $H_0$.[98] To construct the CAP-augmented Hamiltonian, one needs to compute a few eigenstates of the similarity-transformed Hamiltonian, the reduced one-particle density matrix ($\gamma^i$) and transition density matrices ($\gamma^{ij}$) between all the pairs of state i and j. The CAP-matrix elements in the correlated basis are found by transforming the CAP-matrix elements in the atomic orbital basis to the correlated basis (CB) as

$$W_{CB}^{ij} = Tr\left[W^{AO}\gamma^{ij}\right], \qquad i \neq j \tag{22}$$

Hence, the CAP-augmented Hamiltonian becomes,

$$H_{CAP} = H_{CB} - i\eta W_{CB} \tag{23}$$

This scheme enables one to find the optimal $\eta$ value with a single real-values EOM-CC calculation and several diagonalizations of the CAP matrix at different $\eta$ values.

## 3. Computational Details

The neutral geometry of cytosine and microhydrated cytosine complexes were optimized using RI-MP2/def2-TZVP level of theory. Various conformers of the microhydrated cytosine (Cyt(H$_2$O)$_n$, n=1,2,3,4) were generated using CREST[99] software and subsequently optimized using the XTB[100] method. The lowest energy conformer in each category was taken and optimized at the RI-MP2/def2-TZVP level of theory for further calculations. The optimized structures of the isomers are presented in Figure 3. The cartesian coordinates for all the optimized structures are provided in the Supporting Information.

Resonance stabilization curves were generated using the EA-EOM-DLPNO-CCSD[73] method. Firstly, the real part of the energy of several states, which are important for the description of resonances, was calculated by varying the radial extent of the few most diffuse functions. The calculations were performed using the aug-cc-pVXZ (X=D,T,Q) basis set, augmented with additional 1s, 1p and 1d functions added to the heavy atoms (except H-atom) in an even-tempered manner.[101] A factor of 0.5 has been used for generating successive exponents. We have represented these basis sets as aug-cc-pVXZ* (X=D,T,Q) in rest of the manuscript. For the generation of the stabilization plot, the exponent of the two most diffuse s, p and d functions were scaled. The corresponding auxiliary basis set was generated using autoaux[102] utility of ORCA. The full stabilization graph for cytosine plotted using EA-EOM-DLPNO-CCSD method as a function of the scaling coordinate α is shown in Figure 4. The lower panel in Figure 4 demonstrates the stable region corresponding to various shape resonance states of cytosine. Stabilization graphs of all microhydrated cytosine complexes (Cyt(H$_2$O)$_n$, n=1,2,3,4) at different basis sets and truncation level are also given in Supplementary Information. All the calculations were performed with the development version of ORCA,[103] except the projected CAP calculations, which are performed using QCHEM.[104]

Resonance via Padé (RVP) method[65,66] is used to calculate the resonance position and width from the stable region of stabilization plot. Although there might be numerous such branches for each resonance state, the reported resonance energy and width are averages of the statistically best-behaving branches. Additionally, we have computed the resonance states of isolated cytosine using projected-CAP-EA-EOM-CCSD method with box and Voronoi potential using aug-cc-pVDZ* basis set. The onset for the box-CAP is 12.06, 9.07, 3.21 Å in the X, Y, and Z directions and the Voronoi CAP cutoff is 4.18 Å. The resonance states of the monohydrated cytosine were also computed in a similar fashion with projected Voronoi-CAP.

## 4. Results and Discussion
### 4.1. Isolated Cytosine

The focus of this study is to explore the impact of microsolvation on different shape resonance states of cytosine nucleobase, using the EA-EOM-DLPNO-CCSD method. To investigate the accuracy of EA-EOM-DLPNO-CCSD method for the position and width of resonance, we calculated the shape resonance corresponding to 1π*, 2π*, and 3π* states and compared them with existing theoretical and experimental values in the literature. The aug-cc-pVDZ basis set has been used for the calculations with an additional 1s1p1d function added to the heavy atoms, following the work of Matsika and co-workers.[30] Figure 4 shows the stabilization curve obtained for cytosine. The stable region between the avoided crossings were used as input for Padé calculation and are marked separately. The reported resonance position and width values are an average of the two statistically best-behaved branches considered for each resonance state, as shown in the lower panel of Figure 4. The natural orbitals calculated at the stabilized region of different branches were plotted to ensure that the points belong to the same resonance state. Table 1 presents the position and width obtained for three lowest-lying resonance states of isolated cytosine, calculated at the EA-EOM-DLPNO-CCSD/aug-cc-pVDZ* level of theory, as well as in projected-CAP based EA-EOM-CCSD approach, along with previously reported theoretical and experimental results.

In the RVP-EA-EOM-DLPNO-CCSD method, the first two shape resonance states of cytosine are obtained at 0.92 eV and 2.37 eV with lifetimes of 94 fs and 6.6 fs, respectively. The third resonance state is even short-lived and has a lifetime of only 3 fs. The resonance positions and widths of these states agree well with the stabilization-based EA-EOM-CCSD results reported by Matsika[30] and co-workers. However, for the third resonance state, the resonance position is overestimated, and the width is underestimated when compared to reference.[30] It is worth noting that accurate computation of the third shape resonance state of DNA nucleobases can be challenging because of possible mixing with the core-excited resonance states.[29]

Our projected CAP based EA-EOM-CCSD calculation using box CAP potential accurately reproduces the position of the 1π* state but overestimates its width significantly in comparison to both RVP results reported here and the stabilization results reported by Matsika and co-workers.[30] Whereas, the width in the CAP-EA-EOM-CCSD method is closer to the SAC-CI results of Sommerfeld[58] and co-workers. The position and width of the second and third resonance state in CAP-EA-EOM-CCSD is also overestimated compared to our RVP results and the calculated width is comparatively closer to the SAC-CI results for $3\pi^*$ state. However, a box-like CAP not always corresponds to the three-dimensional geometry of the molecule and can be difficult to apply for systems that do not efficiently fill up the vacant space.[95] It will be particularly problematic for microsolvated structures. The Voronoi CAP suggested by Ehara and Sommerfeld[96] shares the same symmetry of the molecular system and more flexible than the box shape CAP, therefore, we have performed projected CAP-EA-EOM-CCSD calculations using Voronoi CAP also.[95] The results for all the three states are similar to that obtained using box-shaped CAP except for the width for $3\pi^*$ state.

The resonance position for all three resonance states obtained in our RVP-EA-EOM-DLPNO-CCSD method is overestimated compared to the experimental values. However, the deviation relative to the experimental value decreases when the energy difference between two resonance states is considered, which shows that the gap between two resonance states is well reproduced in our calculations. There is a wide range of resonance positions reported by various theoretical methods (as listed in Table 1), with the position range for $1\pi^*$ state being 0.36-1.7 eV, for $2\pi^*$, 1.75-4.3 eV, and for $3\pi^*$ goes to 5.35-8.1 eV. Nonetheless, the resonance positions predicted by various EA-EOM-CCSD[30] methods, including our present work, and the SAC-CI method,[58] are generally in reasonable agreement with each other. The decay width or lifetime of the resonance state has a higher degree of uncertainty and shows a larger spread among the methods cited in Table 1.

### 4.2. Impact of DLPNO approximation:

As described previously, the accuracy of the EA-EOM-DLPNO-CCSD method depends mainly upon four truncation thresholds called TCutDo, TCutMKN, TCutPNO, and TCutPairs. This method has demonstrated remarkable success in describing bound state anions. However, this is the first instance where the natural orbital-based approximation to EA-EOM-CCSD has been employed for resonance calculation. Therefore, it is crucial to gauge the accuracy of the EA-EOM-DLPNO-CCSD approach with respect to the truncation parameters. Neese and co-workers[93] have defined a series of composite thresholds named LOOSEPNO, NORMALPNO, and TIGHTPNO, which offer increasingly accurate outcomes at the expense of higher computational costs. Table 2 shows the corresponding values of the various truncation parameters used for these composite thresholds.

Table 3 presents the resonance position and width calculated using RVP-EA-EOM-DLPNO-CCSD/aug-cc-pVDZ* level of theory at LOOSEPNO, NORMALPNO and TIGHTPNO settings. The associated stabilization graphs are given in Supporting Information. The results are compared with the canonical RVP-EA-EOM-CCSD method without any approximation. It is evident that the TIGHTPNO setting shows excellent agreement with the canonical results for all three shape resonance states, with a maximum deviation obtained for the $3\pi^*$ state, where the position is overestimated and the width is underestimated by 0.04 eV. The NORMALPNO setting, on the other hand, offers reasonable agreement only for the resonance position, with the $3\pi^*$ state overestimated by 0.14 eV. The resonance decay widths are underestimated compared to the canonical values. The LOOSEPNO setting results in significantly larger errors for both the resonance position and decay width, with even the $1\pi^*$ state's position being underestimated by 0.16 eV. Therefore, the EA-EOM-DLPNO-CCSD method with TIGHTPNO setting offers the best compromise between efficiency and computational cost for larger calculations.

### 4.3. Impact of aqueous media:

To investigate the impact of aqueous environment on the resonance states of cytosine, four microsolvated cytosine complexes (Cyt(H$_2$O)$_n$ (n=1,2,3,4)) were examined. The presence of water molecules did not alter the qualitative nature of all three resonance states, as evidenced by the corresponding natural orbital provided in Figure 5. However, a part of the additional electron density spilled over to the water molecules in the case of Cyt(H2O)$_4$. The resonance position of all three states experienced a redshift in the presence of water molecules, with an increase in the resonance states' lifetime as the number of water molecules are increased, as shown in Table 4. The magnitude of the redshift in the resonance position varied for the different resonance states. The addition of the first two water molecules resulted in a redshift of 0.07 eV and 0.15 eV for the $1\pi^*$ state, as compared to isolated cytosine. The resonance width also decreased slightly relative to isolated cytosine. Similar results were obtained using the projected CAP-based EA-EOM-CCSD method, where the addition of the first two water molecules led to a redshift of 0.08 and 0.13 eV (see Table S1) in the resonance position for the $1\pi^*$ state.

Further, the addition of third and fourth water molecule in RVP approach resulted in a sharp decrease of 0.19 eV and 0.26 eV, respectively, in the resonance position of the $1\pi^*$ state. In contrast, the second state experienced a decrease of 0.26 eV even with the addition of the first water molecule, demonstrating the stronger stabilizing effect of water molecules for the $2\pi^*$ state. The total stabilization caused by four water molecules for the $2\pi^*$ state is 0.73 eV. The addition of water molecules resulted in even stronger stabilization for the $3\pi^*$ state, where the total redshift caused by four water molecules was 1.16 eV.

With the addition of four water molecules, the first resonance state of cytosine experiences a significant increase in lifetime, from 94 to 658 fs. However, the second and third resonance states show much smaller increases in the lifetime, from 6.6 fs to 33 fs and 3 fs to 4.4 fs, respectively. This suggests that the $1\pi^*$ state, which has a lifetime of 658 fs in the Cyt(H$_2$O)$_4$ molecule, may have enough time in an aqueous solution to convert into a stable valence-bound state, as the time required to form nucleobase-bound anions from pre-solvated electrons falls within the timescale of 500 to 1500 fs.[105] This finding aligns with the recent experimental observation by Verlet[77] and colleagues, who were unable to detect the $1\pi^*$ state in their photoelectron spectroscopy of microsolvated uracil clusters.

It is important to note that conclusions drawn from microsolvation studies may not always be applicable to the bulk solvent environment. Implicit solvent models[106] offer a convenient and black box way to incorporate the effect of bulk solvent in quantum chemical calculations, but recent work by Simons and co-workers[107] has demonstrated that these models are not adequate for describing aqueous anions, even when combined with a microsolvated model. As an alternative approach, explicit solvent models based on QM/MM can be used to simulate the effect of bulk solvation in quantum chemical calculations. To assess the suitability of QM/MM models for simulating resonance states of solvated nucleobases, calculations were performed on Cyt(H$_2$O)$_n$ (n=1,2,3,4) molecules using water as a TIP3P point charge. The results show that the TIP3P model can reproduce the position of the first resonance state accurately, similar to that reported by Matsika and co-workers[108] for monohydrated uracil using QM/EFP calculations. The resonance width in QM/MM also shows reasonable agreement with full QM calculations for the $1\pi^*$ state. However, the QM/MM method does not reproduce the resonance position for the $2\pi^*$ and $3\pi^*$ states as accurately. The inclusion of four water molecules as point charge results in a redshift of 0.49 eV and 0.59 eV for the $2\pi^*$ and $3\pi^*$ states, respectively, compared to the isolated cytosine. Whereas the total shift observed in full QM calculations is 0.73 eV and 1.16 eV for the $2\pi^*$ and $3\pi^*$ states, respectively. Nonetheless, the resonance width for all three states is qualitatively reproduced in the QM/MM studies.

Additionally, to gain a better understanding of the source of stabilization resulting from the explicit water molecules, further calculations were conducted with the water molecules treated as ghost atoms. Table 4 shows that the shift in resonance position and width for the first resonance state with one ghost water molecule is almost identical to that observed in the explicit presence of one water molecule. This finding suggests that the apparent stabilization of the $1\pi^*$ state by one water molecule in full QM is a finite basis set artifact. However, the addition of more water molecules results in real stabilization, which arises from the interaction between water molecules and cytosine. In contrast, for the $2\pi^*$ and $3\pi^*$ states, one water molecule induces physical stabilization in addition to the finite basis set effect. Notably, for all three states, the finite size of the basis set can significantly impact the accuracy of the calculated resonance position and widths. Thus, it is crucial to take into account the effect of the basis set on the computed resonance position and width.

### 4.4. Impact of basis set:

Bhattacharya and co-workers[83] have demonstrated that achieving a proper balance between polarization and diffuse functions is necessary to obtain accurate resonance position and width using the RVP method. To investigate the basis set dependence of resonance position and width, we conducted calculations using the aug-cc-pVXZ* (X=D, T, Q) series of basis sets. Table 5 illustrates the impact of the basis set size on resonance position and width for both the cytosine nucleobase and monohydrated cytosine molecule. For all three resonance states, the resonance position and decay width decrease with the increase in basis set size, which is consistent with the fact that a larger basis set leads to preferential stabilization of the anionic state.

On transitioning from the aug-cc-pVDZ* to the aug-cc-pVTZ* basis set for a cytosine molecule, the $1\pi^*$ resonance position undergoes a redshift of 0.16 eV. However, this shift is notably smaller (0.06 eV) when moving from the aug-cc-pVTZ* to the aug-cc-pVQZ* level. As the size of the basis set increases, the calculated results tend to move towards the experimental results. Nonetheless, even at the aug-cc-pVQZ* level, there is a deviation of 0.38 eV between the calculated $1\pi^*$ resonance position obtained using the RVP-EA-EOM-DLPNO-CCSD method (0.7 eV) and the experimental value (0.32 eV). Similarly, the $2\pi^*$ resonance state experiences stabilization, with the resonance position redshifting by 0.09 and 0.08 eV as the basis set increases from aug-cc-pVDZ* to aug-cc-pVTZ* and then to aug-cc-pVQZ*.

The impact of the basis set is particularly noticeable for the third resonance state. When transitioning from the aug-cc-pVDZ* to the aug-cc-pVTZ* basis set, the resonance position of the $3\pi^*$ state undergoes a redshift of 0.81 eV. However, the shift is substantially smaller (0.19 eV) from the aug-cc-pVTZ* to the aug-cc-pVQZ* level. Hence, one can assume that the aug-cc-pVQZ* basis set offers sufficient accuracy for calculating resonance positions. Similar patterns are observed for monohydrated cytosine, and the basis set correction appears to be almost additive, at least for the first two states. Therefore, one can utilize the Δ basis set correction calculated from the monohydrated cytosine to estimate the basis set corrected value for the Cyt($H_2O$)$_4$ system. This leads to resonance positions of 0.04 eV, 1.85 eV, and 3.84 eV, respectively, for the $1\pi^*$, $2\pi^*$, and $3\pi^*$ states in Cyt($H_2O$)$_4$ complex. Thus, it is evident that, in the presence of bulk water, the first resonance state will transform into a bound state and may not cause bond breakage in DNA strands.

The decay width of the resonance states exhibits a similar trend. At the aug-cc-pVDZ* basis set level, the first resonance state of cytosine has a width of 0.007 eV. However, as one moves to the aug-cc-pVTZ* and aug-cc-pVQZ* basis set levels, the width of the first resonance state ($1\pi^*$) reduces by 0.002 eV and 0.003 eV, respectively. The width of the next two higher energy resonance states ($2\pi^*$ and $3\pi^*$) also decreases with an increase in basis set. The width of the second resonance state decreases by 0.04 eV at both the aug-cc-pVTZ* and aug-cc-pVQZ* levels, while the third state shows a relatively larger change of 0.09 eV when transitioning from the aug-cc-pVDZ* to the aug-cc-pVQZ* basis set. However, the decay width or lifetime of the resonance state is a more sensitive property than the resonance position. Thus, caution must be exercised when comparing the width corresponding to different basis sets.

## 5. Conclusions

The EA-EOM-DLPNO-CCSD method gives excellent agreement with the standard canonical EA-EOM-CCSD results for the shape resonance states of cytosine at a fraction of the computational cost. The presence of a few water molecules does not change the qualitative nature of the three lowest-lying shape resonance states of cytosine. However, the position and the lifetime get significantly affected. The resonance position for all three resonance states undergoes redshift accompanied by an increase in the lifetime with successive addition of water molecules. The 1π* resonance state of isolated cytosine gets converted to an almost bound state in the presence of four water molecules and is unlikely to cause dissociative electron attachment. The accuracy of calculated resonance positions and widths is highly dependent on the size of the basis set used. The standard non-polarizable QM/MM methods do not give consistent performance for all the resonance states for solvated nucleobases. Therefore, further research is required to better understand the resonance states of DNA in the presence of bulk water. Work is in progress in that direction.

## Supporting Information

Cartesian coordinates of the optimized neutral geometry of cytosine and $Cyt(H_2O)_n$ (n=1,2,3,4) complexes; comparison of CAP-EA-EOM-CCSD and RVP-EA-EOM-DLPNO-CCSD results for cytosine and monohydrated cytosine; the employed basis sets and associated energy stabilization graphs at various DLPNO truncation settings; $0^{th}$ order $\eta$ trajectories of resonance states of cytosine and ($Cyt(H_2O)_n$, n=1,2) complexes at CAP-EA-EOM-CCSD/aug-cc-pVDZ* (using box-CAP and Voronoi-CAP) are provided in the Supporting Information.

## Acknowledgments

The authors acknowledge the support from the IIT Bombay, IIT Bombay Seed Grant project (Project no. RD/0517-IRCCSH0-040), DST-SERB CRG (Project no. CRG/2022/005672) and Matrix (Project no. MTR/2021/000420) projects, DST-Inspire Faculty Fellowship (Project no. DST/INSPIRE/04/2017/001730), CSIR-India (Project no. 01(3035)/21/EMR-II), ISRO (Project no. RD/0122-ISROC00-004), UGC fellowship for financial support, IIT Bombay super computational facility and C-DAC supercomputing resources (PARAM Smriti, PARAM Brahma) for computational time.

## Notes

The authors declare no conflict of interest.

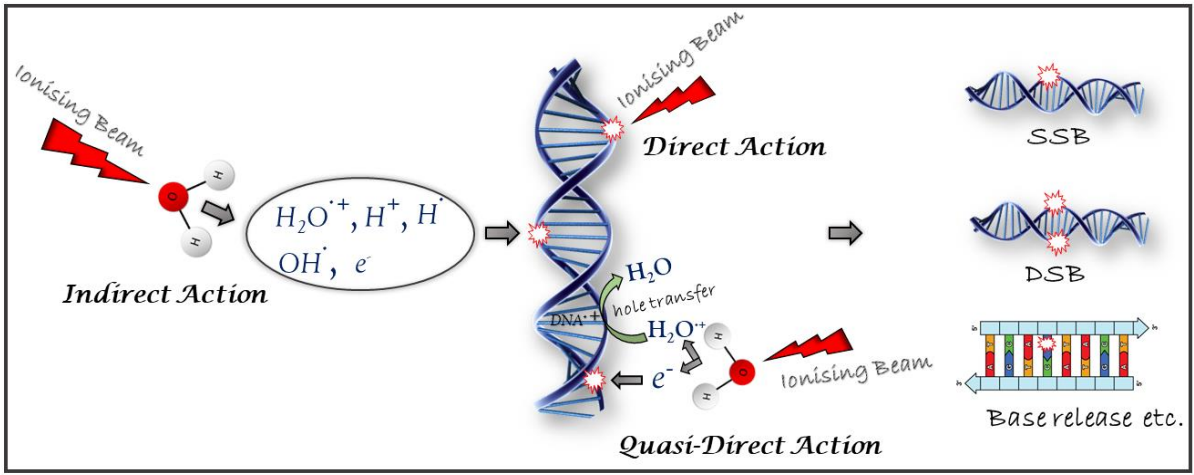

*Figure 1. A schematic representation of the different pathways of radiation damage to DNA.*

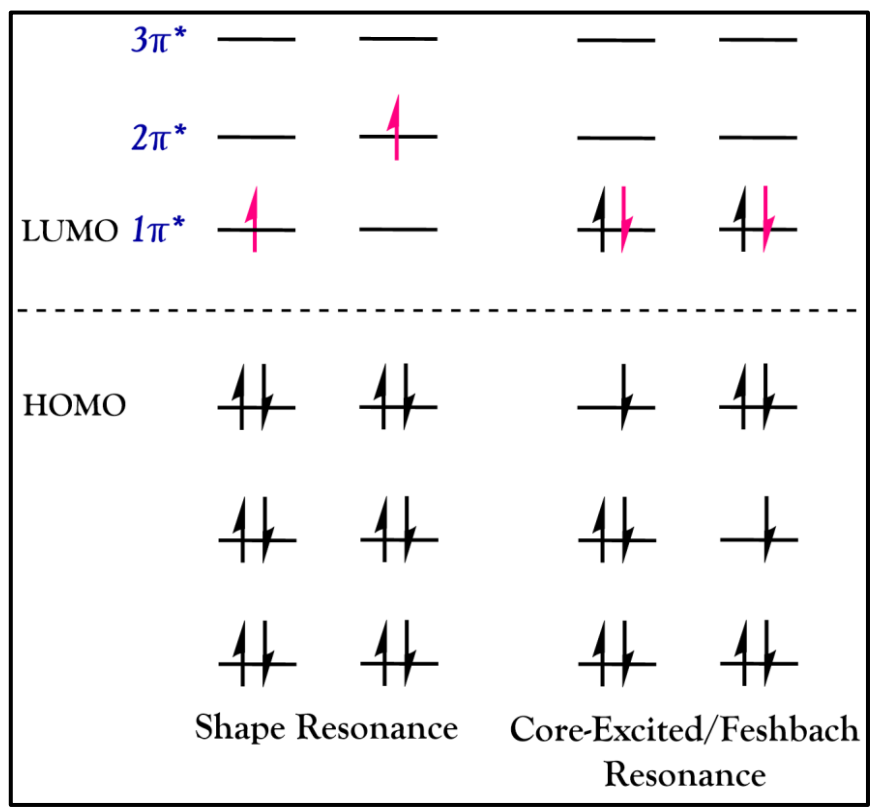

*Figure 2. A schematic representation of the orbital level diagram of shape and core-excited resonance.*

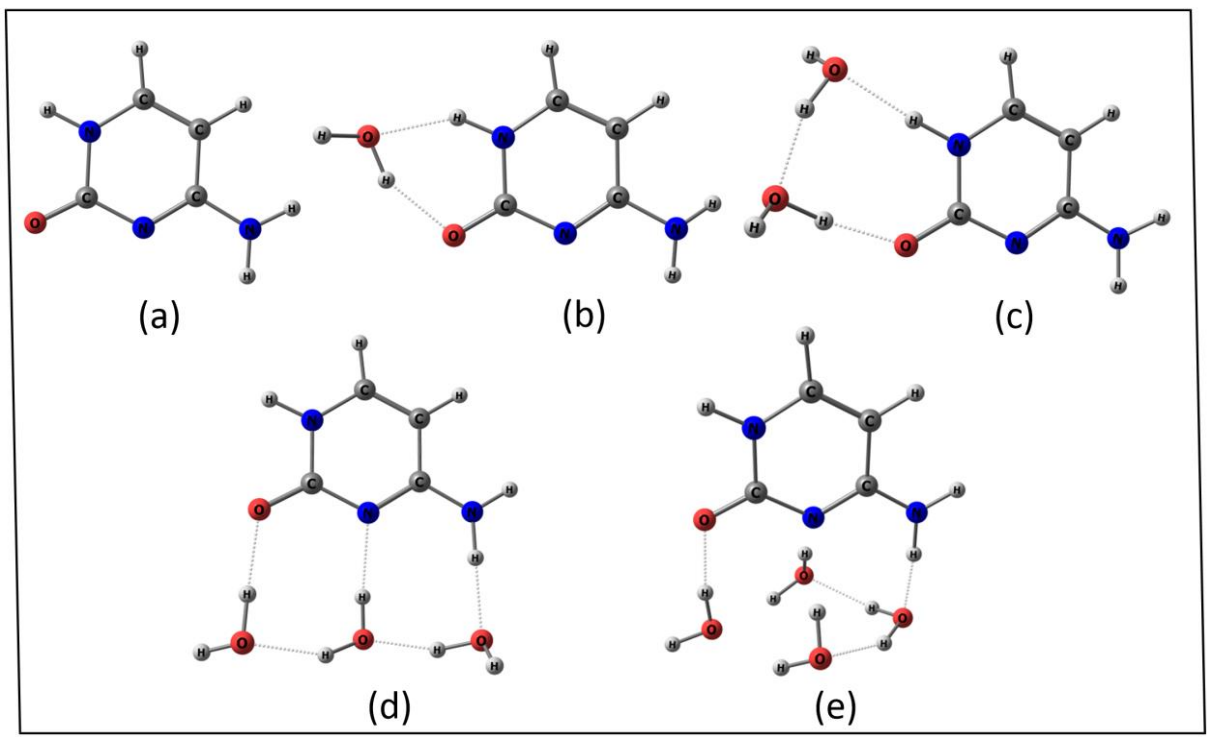

*Figure 3. Optimized structures of cytosine (Cyt) and microhydrated cytosine $(Cyt(H_2O)_n$, n=1,2,3,4) complexes.*

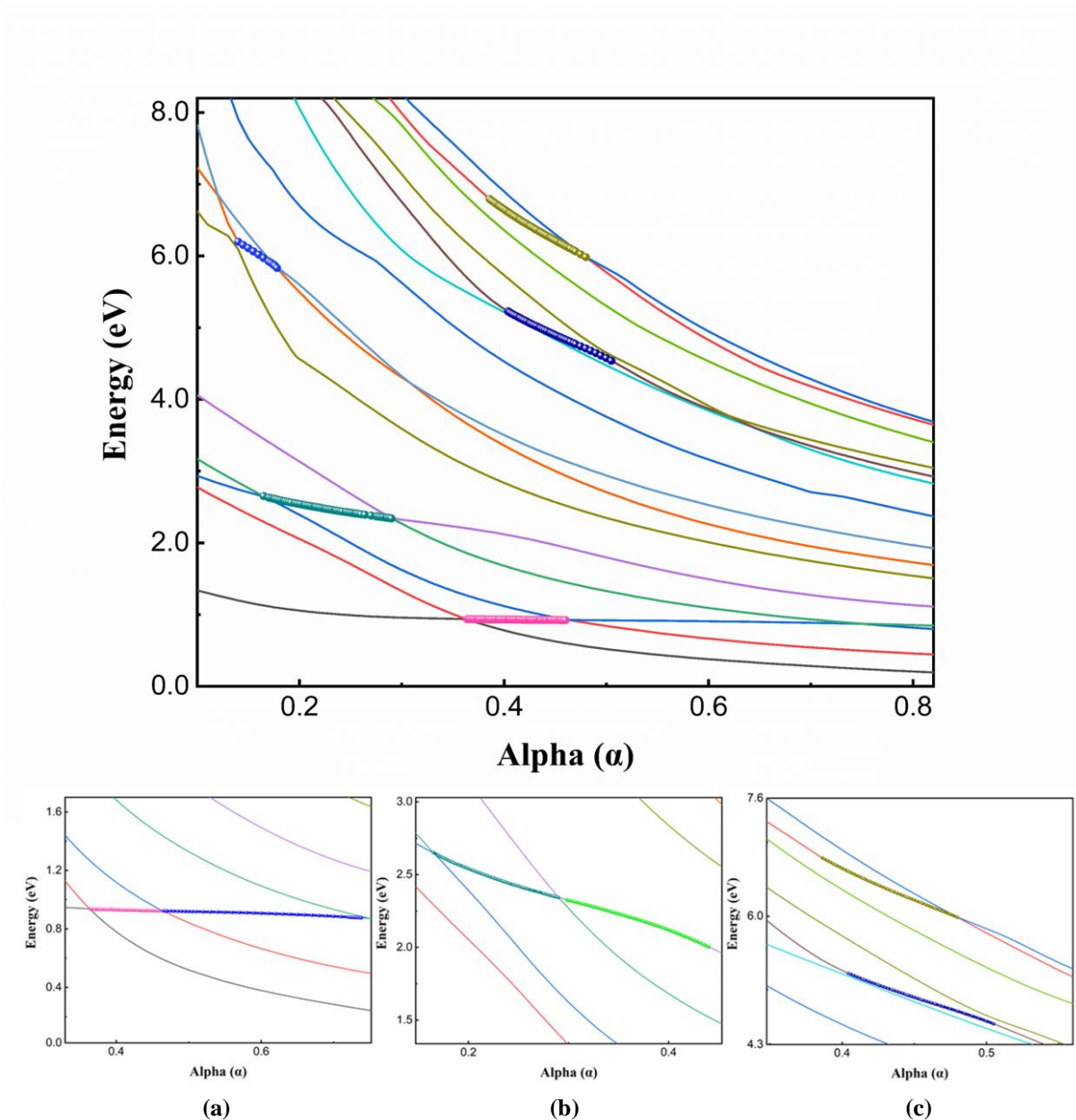

*Figure 4. Stabilization plot of gas phase Cytosine at EA-EOM-DLPNO-CCSD/aug-cc-pVDZ\* level of theory. TIGHTPNO setting has been used. The stabilized region corresponding to (a) 1π\* (b) 2π\* (c) 3π\* shape resonance states of cytosine are separately presented in the below panel. The highlighted part represents the set of points used as the input for RVP method for three resonance states.*

.

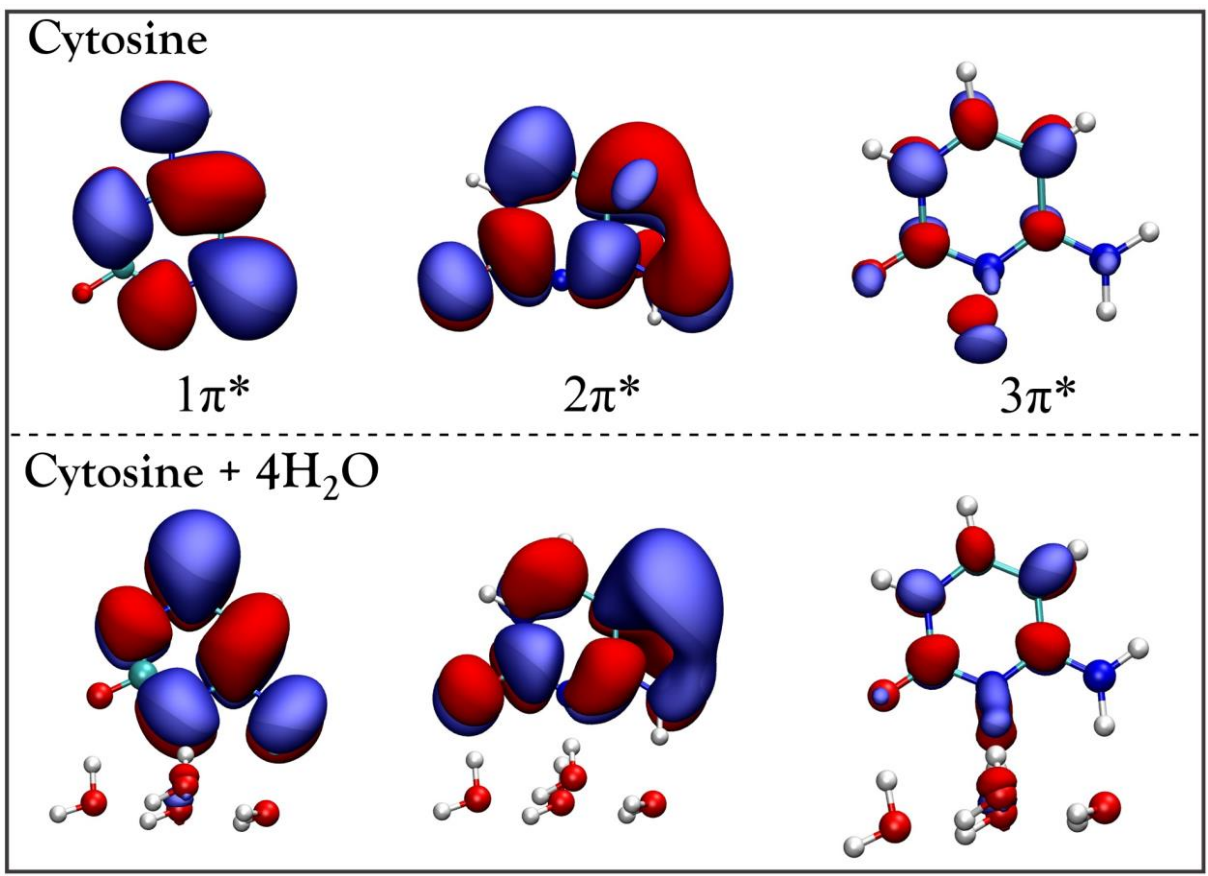

*Figure 5. EA-EOM-DLPNO-CCSD natural orbitals corresponding to (a) 1π\* (b) 2π\* (c) 3π\* shape resonance states of cytosine and Cytosine + 4H$_2$O complex.*

*Table 1. The comparison of resonance position and width for gas phase cytosine at aug-cc-pVDZ\* level of theory.*

| Resonance State | 1π* | | 2π* | | 3π* | |
|---|---|---|---|---|---|---|
| **Method** | $E_R$ | $\Gamma$ | $E_R$ | $\Gamma$ | $E_R$ | $\Gamma$ |
| **EA-EOM-DLPNO-CCSD (TIGHTPNO) (This work)** | 0.92 | 0.007 | 2.37 | 0.104 | 5.72 | 0.221 |
| **Projected CAP (box) EA-EOM-CCSD (This work)** | 0.92 | 0.078 | 2.66 | 0.613 | 5.87 | 0.572 |
| **Projected CAP (Voronoi )EA-EOM-CCSD (This work)** | 0.93 | 0.083 | 2.70 | 0.659 | 5.87 | 0.215 |
| **EOM-EA-CCSD**[30] | 0.93 | 0.017 | 2.40 | 0.19 | 5.54 | 0.35 |
| **CAP /SAC-CI**[58] | 0.7 | 0.16 | 2.18 | 0.30 | 5.66 | 0.63 |
| **R-matrix/u-CC (2012)**[80] | 0.36 | 0.016 | 2.05 | 0.30 | 5.35 | |
| **R-matrix/SEP (2012)**[80] | 0.71 | 0.05 | 2.66 | 0.33 | 6.29 | 0.72 |
| **SMC/SEP (2007)**[109] | 0.5 | | 2.40 | | 6.30 | |
| **R-matrix/SE (2006)**[110] | 1.7 | 0.5 | 4.3 | 0.7 | 8.1 | 0.8 |
| **SMC**[111] | 0.61 | 0.24 | 1.74 | 0.66 | 5.5 | |
| **expt**[44] | 0.32 | | 1.53 | | 4.50 | |

*Table 2. The truncation threshold corresponding to the composite truncation parameters used in the EA-EOM-DLPNO-CCSD calculations.*

| Keywords | TCutpairs | TCutPNO | TCutDO | TCutMKN |
|---|---|---|---|---|
| **LOOSEPNO** | $10^{-3}$ | $10^{-6}$ | $2\times10^{-2}$ | $10^{-3}$ |
| **NORMALPNO** | $10^{-4}$ | $3.3\times10^{-7}$ | $10^{-2}$ | $10^{-3}$ |
| **TIGHTPNO** | $10^{-5}$ | $10^{-7}$ | $5\times10^{-3}$ | $10^{-4}$ |

*Table 3. The effect of truncation parameter on the resonance position and width calculated using EA-EOM-DLPNO-CCSD/aug-cc-pVDZ\* level of theory.*

| Molecule | Canonical | | TIGHTPNO | | NORMALPNO | | LOOSEPNO | |
|---|---|---|---|---|---|---|---|---|
| Cytosine | $E_R$ | $\Gamma$ | $E_R$ | $\Gamma$ | $E_R$ | $\Gamma$ | $E_R$ | $\Gamma$ |
| **1st Resonance** | 0.92 | 0.015 | 0.92 | 0.007 | 0.90 | 0.0035 | 0.76 | 0.0025 |
| **2nd Resonance** | 2.40 | 0.124 | 2.37 | 0.104 | 2.33 | 0.03 | 2.30 | 0.032 |
| **3rd Resonance** | 5.68 | 0.262 | 5.72 | 0.221 | 5.82 | 0.183 | 5.80 | 0.192 |

*Table 4. The effect of microsolvation on the resonance position and width of 1π*, 2π* and 3π* shape resonance states calculated using RVP-EA-EOM-DLPNO-CCSD.*

| Molecule | Full QM | | Water as Point Charge (QM/M) | | Water as Ghost Atom | |
|---|---|---|---|---|---|---|
| | $E_R$ | $\Gamma$ | $E_R$ | $\Gamma$ | $E_R$ | $\Gamma$ |
| **Cytosine** | 0.92 | 0.007 | | | | |
| | 2.37 | 0.104 | | | | |
| | 5.72 | 0.221 | | | | |
| **Cytosine + H$_2$O** | 0.85 | 0.005 | 0.86 | 0.007 | 0.85 | 0.004 |
| | 2.11 | 0.062 | 2.21 | 0.025 | 2.24 | 0.021 |
| | 5.20 | 0.201 | 5.42 | 0.191 | 5.70 | 0.192 |
| **Cytosine + 2H$_2$O** | 0.77 | 0.006 | 0.83 | 0.003 | 0.9 | 0.0035 |
| | 2.05 | 0.052 | 2.19 | 0.012 | 2.24 | 0.052 |
| | 4.85 | 0.173 | 5.52 | 0.182 | 5.34 | 0.165 |
| **Cytosine + 3H$_2$O** | 0.58 | 0.003 | 0.63 | 0.007 | 0.86 | 0.002 |
| | 1.98 | 0.033 | 2.17 | 0.041 | 2.31 | 0.024 |
| | 4.78 | 0.171 | 5.39 | 0.181 | 5.45 | 0.172 |
| **Cytosine + 4H$_2$O** | 0.32 | 0.001 | 0.32 | 0.003 | 0.86 | 0.004 |
| | 1.64 | 0.021 | 1.88 | 0.032 | 2.45 | 0.082 |
| | 4.56 | 0.152 | 5.13 | 0.173 | 5.57 | 0.162 |

*Table 5. Impact of the basis set on the resonance position and width calculated using RVP-EA-EOM-DLPNO-CCSD calculations.*

| Molecule | aug-cc-pVDZ* | | aug-cc-pVTZ* | | aug-cc-pVQZ*[a] | |
|---|---|---|---|---|---|---|
| Cytosine | $E_R$ | $\Gamma$ | $E_R$ | $\Gamma$ | $E_R$ | $\Gamma$ |
| 1st Resonance | 0.92 | 0.007 | 0.76 | 0.005 | 0.70 | 0.002 |
| 2nd Resonance | 2.37 | 0.104 | 2.28 | 0.062 | 2.20 | 0.021 |
| 3rd Resonance | 5.72 | 0.221 | 4.91 | 0.202 | 4.72 | 0.133 |
| Cytosine + $H_2O$ | | | | | | |
| 1st Resonance | 0.85 | 0.005 | 0.69 | 0.003 | 0.57 | 0.003 |
| 2nd Resonance | 2.11 | 0.062 | 2.05 | 0.036 | 1.98 | 0.028 |
| 3rd Resonance | 5.20 | 0.201 | 4.65 | 0.062 | 4.58 | 0.042 |

[a]*NORMALPNO setting has been used for calculations with aug-cc-pVQZ* basis set.*